\documentclass[twocolumn,prb,aps,showpacs,footinbib,amsmath,amssymb,superscriptaddress]{revtex4-1}
\usepackage{graphicx}
\usepackage{dcolumn}
\usepackage{color}
\usepackage{bm}
\usepackage{natbib,hyperref}
\usepackage{amsmath,amssymb,amsfonts,bbold}
\usepackage[normalem]{ulem}

\def\BiPd{$\beta$-Bi$_2$Pd }
\begin{document}


\title{Atomic Manipulation of In-gap States on the \BiPd Superconductor}

\author{Cristina Mier}
\altaffiliation{These authors contributed equally}
\affiliation{Centro de F{\'{\i}}sica de Materiales
        CFM/MPC (CSIC-UPV/EHU),  20018 Donostia-San Sebasti\'an, Spain}
\author{Jiyoon Hwang}
\altaffiliation{These authors contributed equally}
\affiliation{Center for Quantum Nanoscience (QNS), Institute for Basic Science (IBS), Seoul 03760, South Korea}
\affiliation{Department of Physics, Ewha Womans University, Seoul 03760, South Korea}
\author{Jinkyung Kim}
\affiliation{Center for Quantum Nanoscience (QNS), Institute for Basic Science (IBS), Seoul 03760, South Korea}
\affiliation{Department of Physics, Ewha Womans University, Seoul 03760, South Korea}
\author{Yujeong Bae}
\affiliation{Center for Quantum Nanoscience (QNS), Institute for Basic Science (IBS), Seoul 03760, South Korea}
\affiliation{Department of Physics, Ewha Womans University, Seoul 03760, South Korea}
\author{Fuyuki Nabeshima}
\affiliation{Department of Basic Science, University of Tokyo, Meguro, Tokyo 153-8902, Japan}
\author{Yoshinori Imai}
\affiliation{Department of Basic Science, University of Tokyo, Meguro, Tokyo 153-8902, Japan}
\affiliation{Department of Physics, Graduate School of Science, Tohoku University, Sendai, Miyagi 980-8578, Japan}
\author{Atsutaka Maeda}
\affiliation{Department of Basic Science, University of Tokyo, Meguro, Tokyo 153-8902, Japan}
\author{Nicol{\'a}s Lorente}
\email{nicolas.lorente@ehu.eus}
\affiliation{Centro de F{\'{\i}}sica de Materiales
        CFM/MPC (CSIC-UPV/EHU),  20018 Donostia-San Sebasti\'an, Spain}
\affiliation{Donostia International Physics Center (DIPC),  20018 Donostia-San Sebasti\'an, Spain}

\author{Andreas Heinrich}
\email{heinrich.andreas@qns.science}
\affiliation{Center for Quantum Nanoscience (QNS), Institute for Basic Science (IBS), Seoul 03760, South Korea}
\affiliation{Department of Physics, Ewha Womans University, Seoul 03760, South Korea}

\author{Deung-Jang Choi}
\email{djchoi@dipc.org}
\affiliation{Centro de F{\'{\i}}sica de Materiales
        CFM/MPC (CSIC-UPV/EHU),  20018 Donostia-San Sebasti\'an, Spain}
\affiliation{Donostia International Physics Center (DIPC),  20018 Donostia-San Sebasti\'an, Spain}
\affiliation{Ikerbasque, Basque Foundation for Science, 48013 Bilbao, Spain}

\begin{abstract}
Electronic states in the gap of a superconductor inherit intriguing many-body properties from the superconductor.
Here, we create  these in-gap states by
manipulating Cr atomic chains on the \BiPd superconductor. We find that the topological properties of the in-gap
states can greatly vary depending on the crafted spin chain.
These systems make an ideal platform for non-trivial topological phases because of the large
atom-superconductor interactions and the existence of a large
Rashba coupling at the Bi-terminated surface. 
We study two spin chains, one with atoms two-lattice-parameter apart
and one with square-root-of-two lattice parameters.
Of these, only the second one is in a topologically non-trivial phase,
in correspondence with the spin interactions for this geometry.
\end{abstract}

\date{\today}


\maketitle

\section{Introduction}
The scanning tunneling microscope (STM) permits unprecedented
control at the atomic level~\cite{avouris}.  Since the early days
of STM, atoms have been moved, unveiling matter on the atomic
scale~\cite{eigler,joe,meyer,morgenstern,dimas}.  Atoms involve
interactions that can have a profound impact on the electronic
properties of host substrates, as such, designing atomic structures
can lead to creating new quantum states~\cite{khajetoorians_2019}.
Magnetic atoms strongly modify the low-energy electronic
properties of superconductors.  This is due to the appearance of
\textit{in-gap} states caused by the weakening of the Cooper-pair
binding. These in-gap states are usually called Yu-Shiba-Rusinov (YSR)
states~\cite{Yu_1965,Shiba_1968,Rusinov_1969b}.  Recently, the interest on
in-gap states has increased due to the suggestion of \textit{topological}
edge states appearing on chains of magnetic impurities on
superconductors~\cite{Beenakker,Yazdani_2013,Pientka2013,Nadj-Perge_2014,Ruby_2015b,Pawlak_2016,Yazdani,Wiesendanger2018,Choi_2019}.
These zero-energy edge states imply the presence of a topological
superconducting phase. The zero-energy edge states are Majorana bound
states (MBS) with non-trivial exchange transformations. Braiding of
MBS is at the core of current proposals regarding topological quantum
computation~\cite{Kitaev,Simon}.

The STM has become a major tool in the study of
MBS~\cite{Nadj-Perge_2014,Ruby_2015b,Pawlak_2016,Yazdani,Wiesendanger2018,Wiesendanger2019,Yazdani2021}.
Indeed, its spectroscopic capabilities render it unique for
revealing in-gap states, granting access to unrivaled energy and
space resolutions. Recently, the spatial distribution of in-gap
states has been shown and used to infer new properties on the states
themselves~\cite{Menard_2015,Ruby2016,Choi2017}.  The fore-mentioned
STM manipulation can be used to create atomically precise spin chains
on superconductors~\cite{Wiesendanger2018,Choi_2019,Kampl}.  The new
in-gap states evolve into bands and open gaps displaying new forms of
superconductivity~\cite{Beenakker,Yazdani_2013,Pientka2013,Peng_2015}.
This proves the complexity of the induced electronic structure.
Each added impurity locally creates a few states in the superconducting
gap. As the number of impurities grows, the gap \textit{fills} up with
new quasiparticle states.

The study of impurity dimers illustrates the initial steps of in-gap bands
~\cite{FlatteReynolds,Liljeroth_2018,Ruby_2018,Choi_2018,Beck_2020,Ding2021}.
The quasiparticle states themselves are difficult to describe. In
the Bogolioubov-de Gennes approximation, the quasiparticle states
are taken as electron and hole superpositions despite violating
particle-number conservation. Furthermore, the quasiparticle states are
spin polarized~\cite{Wiebe,Wiebe_2021_a}, which has important implications
in the way the in-gap states hybridized~\cite{FlatteReynolds}. In
particular, the resulting states reflect the spin-ordering of the
magnetic impurities~\cite{Choi_2018}. However, recent work suggests that
in the presence of strong Rashba coupling, it is difficult to conclude
on the actual spin orientation of the impurities by studying the in-gap
states~\cite{Beck_2020}.

Here, we study atomic spin chains of Cr adsorbed on the hollow
sites of \BiPd and  grown along the two main surface directions, the
$\langle100\rangle$ and $\langle110\rangle$ for the Bi-terminated $[001]$
surface using a home-built dilution fridge STM~\cite{kim2021spin}.
By doing so, we are choosing two starkly different spin orientations
for the chain ground state as concluded in Ref.~\onlinecite{Choi_2018}.
Dimers along the $\langle100\rangle$-direction with a Cr-Cr
distance of two-unit cells  ($2a$, where $a$ is the surface
lattice parameter) present antiferromagnetic (AFM) coupling of
their 5$\mu_B$ magnetic moments~\cite{Choi_2018}.  Dimers along the
$\langle110\rangle$-direction are $\sqrt{2}a$ apart, and they are
instead ferromagnetically (FM) coupled.  Here, we compare dimers,
trimers and tetramers of these two types of chains and conclude that the
$\sqrt{2}a-\langle110\rangle$ chains are indeed FM coupled by comparing
with model calculations of spin chains solving the Bogolioubov-de Gennes
equations~\cite{FlatteReynolds,Choi_2018}.  As clearly seen in this work,
the gap closes rapidly for the $\sqrt{2}a-\langle110\rangle$ chains,
however the $2a-\langle100\rangle$ chains maintain an almost constant gap
for chains as long as 12 Cr atoms. This has important implications for the
possibility of engineering topological phases on the \BiPd superconductor.

\begin{figure*}[t]
\begin{center}
\includegraphics[width=0.6\textwidth]{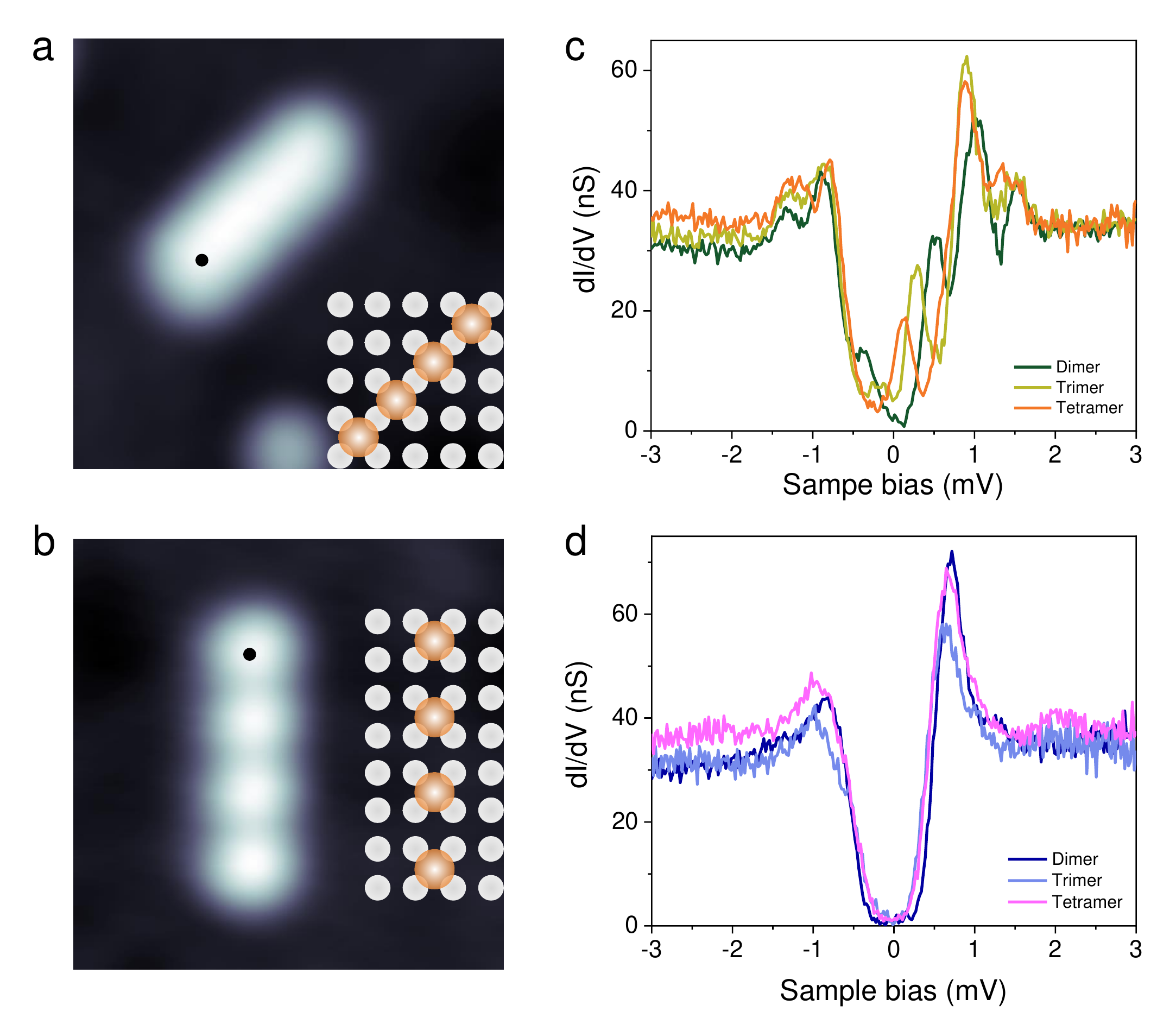}
\end{center}
        \vspace{-0.3cm}
\caption{
Chromium  chains built on \BiPd surface by atomic manipulation.
Topographic images of (a) $\sqrt{2}a-\langle110\rangle$  and (b) $2a-\langle100\rangle$
unit-cell apart tetramer chains ($100$ mV, $10$ pA, $4\times4$
nm$^2$). The insets show the atomic geometry of the tetramer nanostructures.
The corresponding differential conductance is measured at the end atom (marked
black dot) from dimer, trimer to tetramer in (c) $\sqrt{2}$ unit-cell
apart and (d) $2$ unit-cell apart tetramer chains. ($T=30$ mK, AC modulation
bias $30~\mu V$)
 \label{figure1}
}
\end{figure*}

\section{Methods}
\subsection{Sample preparation and STM/STS characterization} 

The \BiPd crystal was fabricated by the method written in Ref.~\onlinecite{imai_2012}.  
The chosen sample showed a $T_c$ of 5.2 K.
The Bi-terminated surface of the \BiPd crystal was prepared by exfoliation
in-situ~\cite{Choi_2018}.  Cr atoms were deposited onto a precooled \BiPd
surface at a temperature  $T \leq 20 $ K to have single isolated atoms.
The experimental data were taken  using a home-built dilution fridge STM 
at $T=30$ mK and in ultra-high vacuum 
at the IBS Center for Quantum Nanoscience~\cite{kim2021spin}.  
The very-low temperature leads to
a negligible thermal smearing granting a resolution higher than the one
obtained by a superconducting tip~\cite{Rodrigo,QKX,Mier2021}. We used a metallic
PtIr tip that permitted us to use the differential conductance, $dI/dV$,
as a direct measurement of the density of states of the substrate
(refer to the supplementary materials\cite{SI} for more details).
The conductance was measured using a lock-in amplifier with AC modulation
bias $30~\mu V$ and frequency $330~Hz$. 

Lateral atomic manipulation
was achieved by approaching the STM tip to one side of a selected
atom to reach junction resistances in the order of a few tens of k$\Omega$
(typically 3 mV and tens of nA). Then the STM tip was laterally moved
to drag the atom to a desired position with the feedback loop open.

\subsection{Theory}
We model the Cr spin chain in the dilute spin chain
limit~\cite{Pientka2013} because density-functional-theory (DFT)
 calculations show that no Cr $d$-states
lie at the Fermi energy, preventing charge transfer processes~\cite{Choi_2018}. In
this framework, 
we solve a spin-chain using  Green's functions for the superconductor in the Nambu basis
set~\cite{Flatte_2000,Flatte_1997}.  We add a Rashba term to the
Hamiltonian expressed in the local basis. The resulting density of states
corresponds to the Bogolioubov-de Gennes states of a BCS superconductor
in the presence of an array of classical spins and subject to the strong
Rashba interaction of the Bi-terminated surface. 

The Fermi velocity entering the superconductor's Green
function~\cite{Flatte_2000,Flatte_1997} is taken to be 0.15 (Hartree atomic
units $\hbar=m=e=1$) and the Dynes parameter~\cite{Dynes} controlling the width
of the superconducting quasiparticle peaks
is 0.05 meV. This leads to peaks in the density of states (DOS) sharper
than the experimental ones, but helps the visualization
of the evolution of in-gap states with the number of Cr atoms.
\BiPd is an s-wave superconductor that can be well accounted for by
a single gap~\cite{imai_2012,Herrera}  $\Delta=0.76$~meV.  
For the normal metal DOS, we use $N=0.037/$eV
that is 5 times larger than the corresponding $N$ for a free-electron
metal with Fermi velocity
of 0.15 atomic units, in order to capture the 5 electrons
of the Bi valence shell.
The Hamiltonian taking into account the superconductor is:
\begin{equation}
    \hat{H}_{BCS} = \xi_k\tau_0\sigma_3 + \Delta \tau_2\sigma_2
    \label{BCS}
\end{equation}
{Where $\sigma_i$ ($\tau_i$) are the Pauli matrices acting on the
spin (particle) sectors. $\xi_k$ is the energy from the Fermi level
($\xi_k = \epsilon_k - E_F$), the previous Hamiltonian is written in the
4-dimensional Nambu basis: $\Psi = (\hat{c}_{\uparrow}, \hat{c}_{\downarrow}, \hat{c}_{\uparrow}^\dagger, \hat{c}_{\downarrow}^\dagger)^T$}.

To model the experimental system, we add the Hamiltonian describing the magnetic impurities~\cite{Flatte_2000,Flatte_1997}.
To do this, we change to a tight-binding basis, assuming a single, very-compact, atomic
orbital per site. Additionally, the interactions with the magnetic impurity is assumed to be strictly localized
to the site where the impurity is sitting~\cite{Pientka2013}. The Hamiltonian is then:
\begin{equation}
 \hat{H}=   \hat{H}_{BCS} + \hat{H}_{impurity} =\hat{H}_{BCS} + \sum_j^N (U_j \tau_3\sigma_0+J_j\vec{S_j}\cdot\vec{\alpha})\; 
\end{equation}
with $\vec{\alpha} = \frac{1 + \tau_3}{2}\vec{\sigma} + \frac{1 -
\tau_3}{2}\sigma_2\vec{\sigma}\sigma_2$, where $\vec{\sigma}$ is the spin
operator~\cite{Shiba_1968}.
This Hamiltonian describes  a BCS superconductor and the interaction between its electrons and $N$ extra impurities.
The interaction contains an exchange coupling, with strength $J_j$, and a non-magnetic potential scattering term, $U_j$,  per impurity $j$. We will use the same impurity species, Cr,
and assume that they are equivalent regardless of their adsorption site and spin chain, in
order to study the system's evolution with the number of atoms in the spin chains.
And $\vec{S_j}=(S_{j,x},S_{j,y},S_{j,z})$ {$= S(\sin{\theta_j}\cos{\phi_j}, \sin{\theta_j}\sin{\phi_j}, \cos{\theta_j})$} is
the spin of atom $j$ considered to be a classical spin and hence not an operator. The local term $U_j$ describes a scalar potential
acting on the substrate's electron. It is responsible for the potential
scattering term produced by the impurity. In the
case of a charged impurity, $U_j$ is mainly given by the Coulomb interaction between
the total charge of the impurity and the charge of the substrate's electron.
 The potential scattering
that explains the electron-hole asymmetry of the YSR bands is taken
as $U_j=5.5$ eV. The values for the Kondo-exchange coupling, $J_j$, are about 2~eV
as estimated from fittings to a single-Cr YSR states~\cite{Choi_2018}.

The Hamiltonian is completed by a Rashha term:
\begin{eqnarray}
\label{Rashba}
    \hat{H}_{Rashba}&=&i \frac{\alpha_R}{2a} 
 \sum_{i,j,\alpha, \beta}[\hat{c}^\dagger_{i+1,j, \alpha}
(\sigma_2)_{\alpha, \beta} \hat{c}_{i,j, \beta} \nonumber \\
&-&\hat{c}^\dagger_{i,j+1, \alpha}
(\sigma_1)_{\alpha, \beta} \hat{c}_{i,j, \beta} +h.c.]
\end{eqnarray}
where ${\alpha, \beta}$ are spin indexes. The interaction
couples spins on nearest-neighbour sites. The lattice parameter of the substrate
is $a$, and the factor of $2 a$ comes from a finite-difference scheme to
obtain the above discretized version of the Rashba interaction. For the case of Bi$_2$Pd, we use a large Rashba
coupling, $\alpha_R\approx 1.8$eV$\cdot$\AA~ as coming from our DFT calculations and in agreement with the couplings of Bi-terminated surfaces~\cite{Bi}.

The local or projected DOS (PDOS) is computed over every local orbital $i$ of the basis using,
\begin{equation}
\rho (i,\omega) = -\frac{1}{\pi} Im  [ 
{G}^{1,1}_{i,i} (\omega) +{G}^{4,4}_{i,i} (-\omega)  ] 
\label{rho}
\end{equation}
{Where ${G}^{\nu,\mu}_{ii}$ is the resulting Green's function evaluated on orbital $i$
for the Nambu components $\nu$ and $\mu$ by solving Dyson's equation:}
\begin{equation}
\label{Dyson2}
    \hat{G} = [\hat{G}_{BCS}^{-1} - \hat{H}_{I}]^{-1}
\end{equation}
Where $\hat{G}_{BCS}$ is the retarded Green's operator for the BCS Hamiltonian from Eq.(\ref{BCS}) and $\hat{H}_I = \hat{H}_{impurity} + \hat{H}_{Rashba}$.

The DFT calculations were performed using the VASP code~\cite{vasp}. The
\BiPd slab was optimized using the Perdew-Burke-Ernzerhof (PBE) form
of the generalised gradient approximation (GGA)~\cite{PBE}, following
the calculations of Ref.~\onlinecite{Choi_2018}. For more details, please
see the supplementary materials~\cite{SI}.

\begin{figure*}[t]
 \begin{center}
		\includegraphics[width=0.9\textwidth]{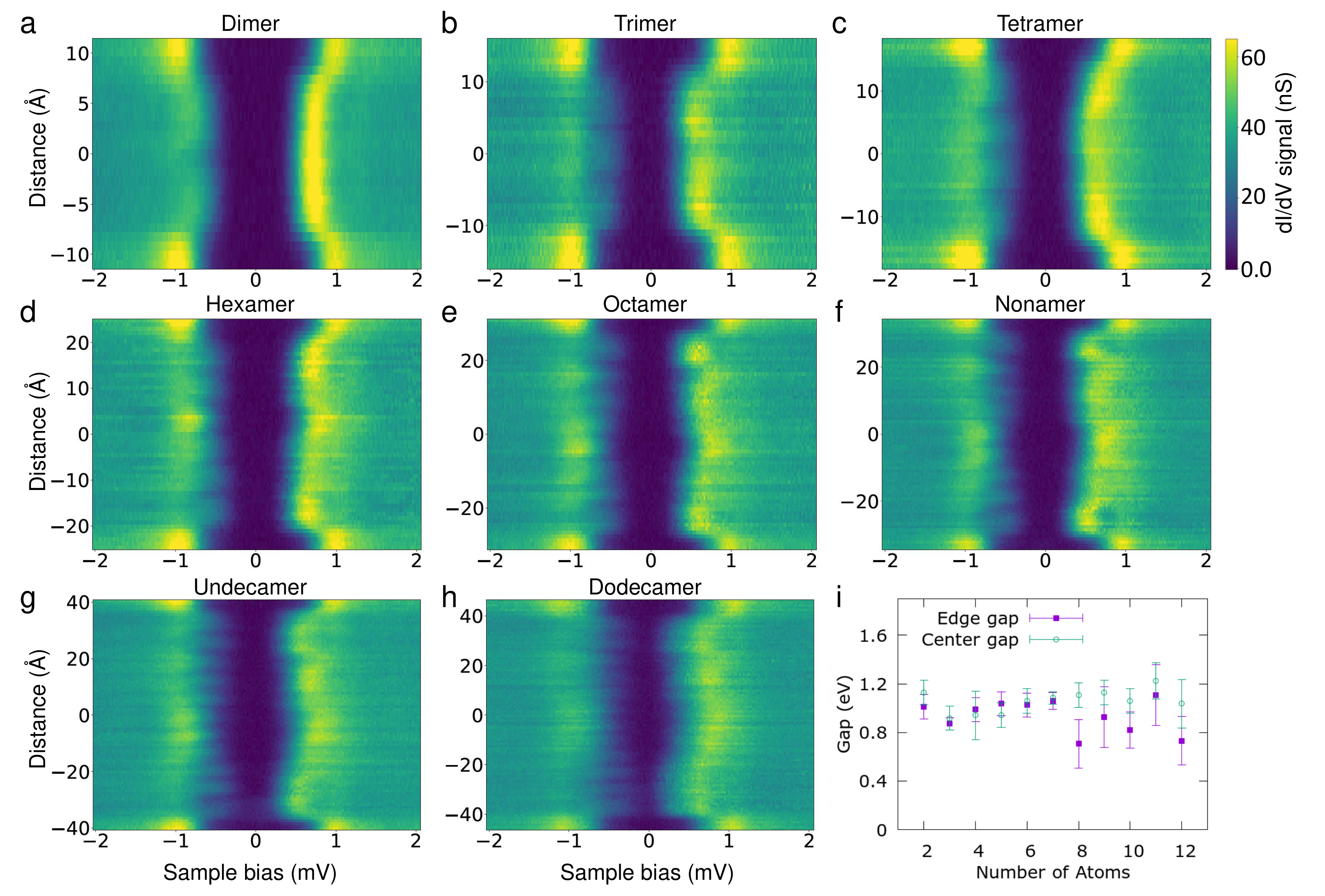}
 \end{center}
	\vspace{-0.2cm}
	\caption{
		Differential conductance measured along
		 Cr$_n$ $2a-\langle100\rangle$ chains
with $n=2$ in (a), $n=3$ in (b), $n=4$ in (c), $n=6$ in (d), $n=8$ in (e),
$n=9$ in (f), $n=11$ in (g), $n=12$ in (h). 
The $x$-axis represents the sample bias, the
$y$-axis displays the distances over the chain. The color code
gives the intensity of the differential conductance. 
The smallest gap in the system, defined as the distance between the
lower quasiparticle peak and the highest quasihole peak is plotted in (i).
In the absence of Cr atoms, the gap corresponds to $2\Delta$ where $\Delta=0.75$ meV
for Bi$_2$Pd. The gap has been obtained at an edge atom or at the center of
the spin chain. 
		\label{figure2}
	} 
\end{figure*}

\begin{figure*}[t]
\begin{center}
		\includegraphics[width=1.0\textwidth]{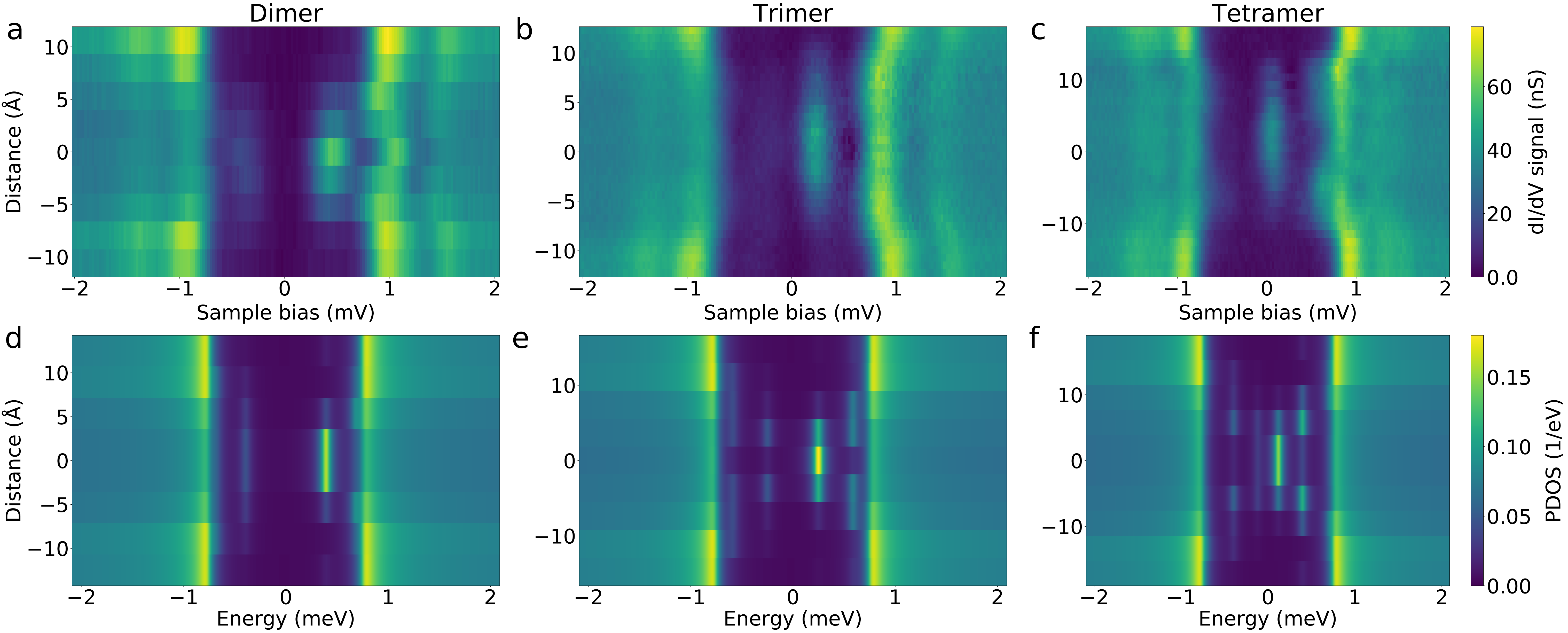}
\end{center}
	\vspace{-0.3cm}
	\caption{
Experimental differential conductance measured along the Cr adatoms of
the  $\sqrt{2}a-\langle110\rangle$	  (a) dimer, (b) trimer and
(c) tetramer chains. 
The corresponding calculations for ferromagnetically coupled $\sqrt{2}a-\langle110\rangle$
(d) dimer, (e) trimer and (f) tetramer chains are depicted in the lower
row.  The color code refers to the projected density of states (PDOS)
on the different sites of the tight-binding lattice, in this case the
one corresponding to the Cr adatoms.  
\label{figure3} } 
\end{figure*}

\section{Results and Discussion}

The  $dI/dV$ over
a  single Cr adatom yields a single YSR state given by  peaks at
$V=\pm 0.35$ mV (please see Refs.~\onlinecite{Choi_2018,SI}).  By lateral
atomic manipulation, we place Cr atoms to create
linear $\sqrt{2}a-\langle110\rangle$ or $2a-\langle100\rangle$
chains.  Figure~\ref{figure1} (a) and (b) show constant-current
images of the two tetramer chains.  The (a) chain corresponds to
the $\sqrt{2}a-\langle110\rangle$ tetramer as depicted in the inset,
the (b) is the $2a-\langle100\rangle$ tetramer. As the chain is made
larger, misplacing a Cr atom becomes more common. Indeed, error-free
$\sqrt{2}a-\langle110\rangle$ spaced nanostructures were difficult to
obtain, while $2a-\langle100\rangle$ chains are easier to manipulate. The
reason lies in the chemistry of the chains. For the more compact chains,
the affinity of Cr atoms for certain conformations 
leads to non-linear arrangements.
The less compact $2a-\langle100\rangle$
chain is easier to fabricate by single-atom manipulation because 
the atoms do not approach each other as much and hence cluster formation is much less common.

Our DFT calculations yield
a coherent picture with the experiment. The Cr atoms are preferentially
adsorbed on the hollow sites of the Bi-riched surface~\cite{Choi_2018},
and the Cr--Cr interactions in the chains are mediated by a single Bi
atom in the $\sqrt{2}a-\langle110\rangle$ chains or a square of four Bi
atoms in the $2a-\langle100\rangle$ chains.  Short $1a-\langle100\rangle$
chains can also be obtained, but the structures easily become clusters
due to the Cr-Cr interaction.  The $\sqrt{2}a-\langle110\rangle$
dimer is 249 meV less stable than the $1a-\langle100\rangle$ dimer. As
a consequence, shifting a single Cr atom towards another Cr to reach
the short $\sqrt{2}a$ distance likely produces a $1a-\langle100\rangle$
dimer. This stacking error becomes more likely as the chain is manipulated
more times to make it longer.  The $2a-\langle100\rangle$ dimer is only 30
meV less stable than the $\sqrt{2}a-\langle110\rangle$.  But, still the interactions
between atoms for the larger Cr--Cr distance,  $2a-\langle100\rangle$
chains, are weaker resulting in an easier manipulation  to
build longer chains. Indeed, the bottom-up approach of chain building is difficult on
many other substrates~\cite{Steinbrecher2017}. 
Recent experiments show long Mn chains built in a similarly compact
geometry as here but on a Nb(110) substrate, also
giving rise to topological in-gap behavior~\cite{Schneider1,Schneider2}.

Once the chains are built, the differential conductance, $dI/dV$, as a
function of bias, $V$, and surface position, is an extraordinary probe
of the electronic properties of the new systems.  Figure~\ref{figure1}
(c) and (d) shows $dI/dV$ spectra measured at $T~=~30$ mK for the
dimer, trimer and tetramer of $\sqrt{2}a-\langle110\rangle$ and
$2a-\langle100\rangle$ type, respectively.  The $dI/dV$ spectra are
taken at an edge atom (black dot in Fig.~\ref{figure1} (a) and (b)).
The two sets (c) and (d) are starkly in contrast. Figure~\ref{figure1}
(c) clearly shows an in-gap state that is shifting towards zero bias as
the chain gets longer. With opposite behavior, Fig.~\ref{figure1} (d)
shows no clear in-gap state and a well-formed gap. Furthermore, the gap
for the dimer is larger but the trimer and tetramer show similar gaps,
pointing at a rapid stabilization of gap with chain size.

The in-gap states of the $\sqrt{2}a-\langle110\rangle$ dimer agree well
with a model of two FM aligned spins  and they disappear  if the magnetic
moments are coupled AFM as in Fig.~\ref{figure1} (d) for a $2a-\langle100\rangle$
dimer~\cite{Choi_2018}.  The $2a-\langle100\rangle$-dimer behavior can
be justified by the mutual cancellation of both spins, although the
actual explanation is more involved.  This can be seen by studying the
spatial distribution of the differential conductance along the two types
of chains. Figure~\ref{figure2} and \ref{figure3} show the $dI/dV$ in a
color code (bright-yellow corresponds to larger conductace and dark-blue
represents zero conductance) along the chain,  $y$-axis in \AA~of the
distances over the chain, and $x$-axis in mV of the STM junction's bias.

\begin{figure*}[t]
	\includegraphics[width=0.9\textwidth]{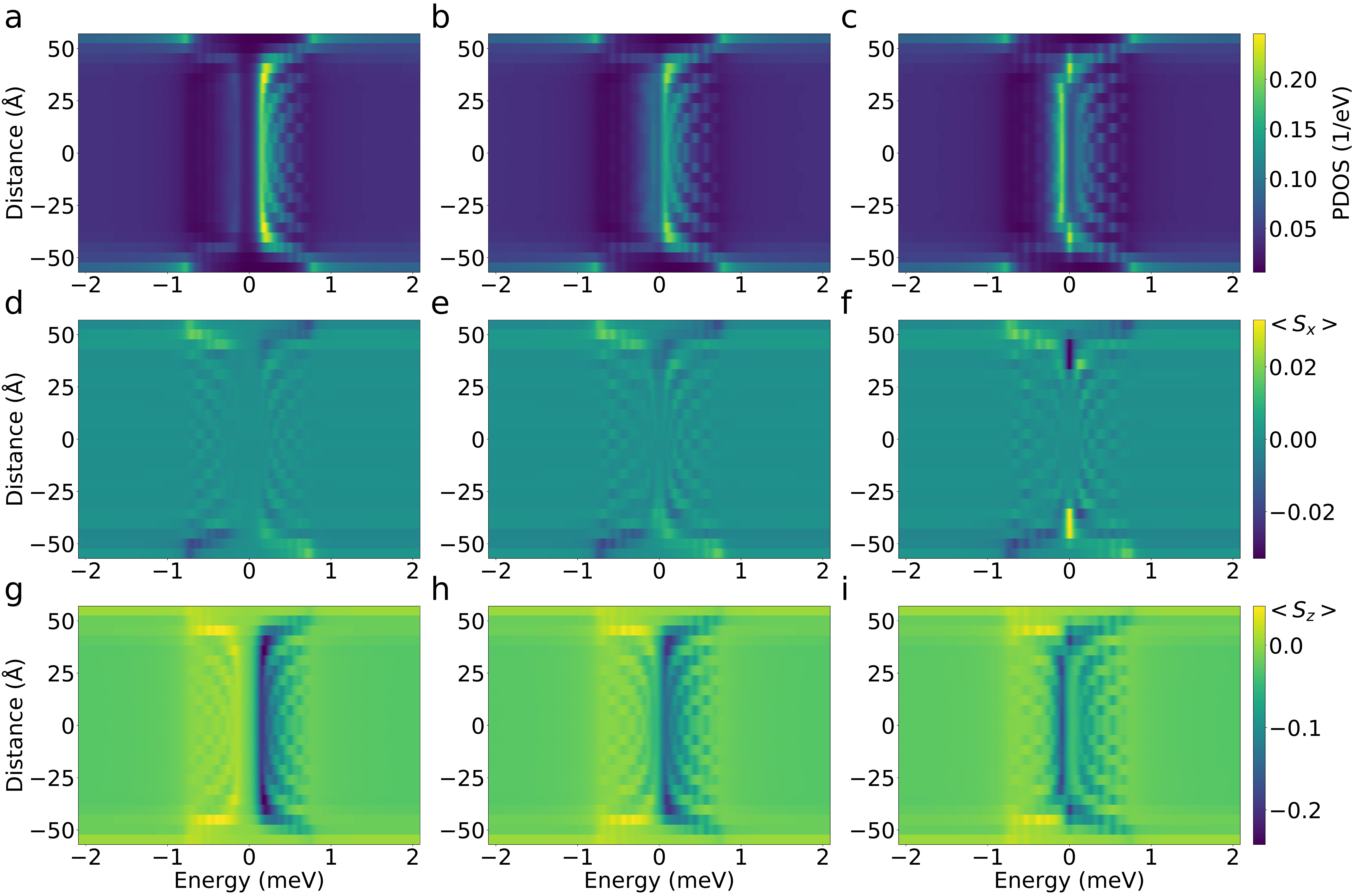}
	\vspace{-0.3cm}
	\caption{
Topological phase transition induced by increasing the exchange
coupling $J$. The three rows of pannels of this figure
correspond to three different values of the exchange
coupling $(a)$ $J=2.1$ eV, $(b)$ $J=2.3$ eV and $(c)$
$J=2.5$ eV for the PDOS showing the quasiparticle states
induced by a Cr$_{20}$ $\sqrt{2}a-\langle 110 \rangle$
chain.	We see that the gap is virtually closed for $(b)$
$J=2.3$ eV and reopens for $(c)$ $J=2.5$ eV displaying
the MBS that indicates the change of topological phase of
the superconductor. Pannels $(d)$, $(e)$, $(f)$ correspond
to the respective values of $J$ and show the transversal
spin density $\langle S_x \rangle$ along the chain.
We see that $\langle S_x  \rangle$ only becomes large and
of opposite sign at the two MBS. Finally, pannels $(g)$,
$(h)$, $(i)$ show the spin density, $\langle S_z \rangle$,
of the YSR states for the three different couplings.
We find that the spin across the gap reverts when the
TPT is achieved and the corresponding MBS have the same
well defined spin.  
\label{figure4} }
\end{figure*}

Figure~\ref{figure2} shows the results for the $2a-\{100\}$ spin
chains. As seen in Fig.~\ref{figure1} (d), we find no obvious
structure in the gap in any of the studied chains.  A closer look reveals
atomic modulations of the quasihole states that match the number of
atoms in the chains. The presence of YSR states can be inferred by the
profile of the gap. The complete sequence of chains from $n=2$ to $n=12$
can be found in the supplemental material~\cite{SI}.  All chains
roughly show a smaller gap at the edge atoms than at the center of the
chain, see Fig.~\ref{figure2} (i). 
In first approximation, the gap is constant with chain length.  Beyond 8
atoms, the chains show a smaller gap at the edge.  However, the closing of
the gap is very small and almost constant for longer chains. These data
indicate that the YSR states are not able to close the superconducting gap,
preventing any topological phase transition. 

Figure~\ref{figure3} presents the $dI/dV$ maps of the
$\sqrt{2}a-\langle110\rangle$ chains (upper row) compared to 
model calculations of the  
PDOS on the surface sites (lower row).
Figure~\ref{figure3} shows excellent agreement between experiment 
and theory if the magnetic moments are FM coupled, which is 
also in good agreement with the results of Ref.~\onlinecite{Choi_2018}.
The calculations of the YSR structure confirm the FM ordering for Cr atoms
sitting along the $\sqrt{2}a-\langle110\rangle$ hollow sites.  Moreover,
the magnetic ordering is not altered by adding extra atoms to the
dimer.

The data of Fig.~\ref{figure3}  permit us to have a clear picture of
the in-gap states for the $\sqrt{2}a-\langle110\rangle$ chains. The
dimer presents two YSR bands, one closer to zero energy with a larger
density of states between the two Cr adatoms, and one closer to the
quasiparticle continuum with a minimum between the atoms.  Adding one
more atom to form the trimer shifts the lowest-energy YSR state closer
to zero, but keeps its overall spatial distribution with a maximum PDOS
on the central atom. Furthermore, we find the second band closer to
the quasiparticle continuum, and again with a minimum of PDOS over the
central point of the chain. We also notice that as in the dimer case,
the quasiparticle PDOS presents a reduction and an oscillation along the
chain. Finally, the tetramer shifts both bands closer to zero, but largely
keeping their spatial distributions. The PDOS at the quasiparticle edge
presents the same features as for the dimer and trimer.

In order to match the very fast experimental closing of the  gap with the chain length, 
the Kondo exchange coupling ($J$) is increased from $J=2.0$ eV for the dimer,
$J=2.1$ eV for the trimer, and $J=2.3$ eV for the tetramer, respectively
in Fig.~\ref{figure3} (d), (e) and (f).  This behavior can be rationalized
by a possible geometrical and electronic rearrangement of the chain as the
spin chain grows in size. The atoms place themselves more symmetrically
and closer to the surface leading to a larger hybridization with the
substrate and thus to larger couplings.

The MBS appear naturally as soon as the exchange coupling $J$ is larger
than 2.3 eV. It is interesting to study how the appearance of MBS takes
place as $J$ varies. This is plotted in Fig.~\ref{figure4}.  The pannels are arranged
in three columns. Each column corresponds to a different value of $J$. The
first one is $J=2.1$ eV, the second one $J=2.3$ eV and the fourth one
is $J=2.5$ eV. The first row plots the PDOS along the chain ($y$-axis) as
a function of the quasiparticle energy ($x$-axis). We see the formation
of YSR bands already for this 20-atom chain. In the middle of the chain,
there is a clear gap in the YSR structure. For small $J$, this gap is
maintained all along the chain, for the larger $J$, the gap is closed
by an edge state that is a MBS as we shall briefly see. For $J=2.3$ eV,
we see that the lowest-energy bands are still separated by a very small
gap, almost closing and for $J=2.5$ the gap is well-formed again. The
closing and reopening of the gap is a necessary condition to change to
a topologically non-trivial superconducting band structure.

The second row is the transversal spin density component $\langle S_x
\rangle$ along the chain for the same YSR state as above. We see that the
values are small and dispersed for $J=2.1$ and $2.3$ eV. For $J=2.3$ eV
the values of $\langle S_x  \rangle$ extend all over the superconducting
gap giving the impression of many YSR states closing the gap. But $J=2.5$ eV
is very different. The gap in $\langle S_x  \rangle$ is again clear and
very sharp values at just the edge states appear and are of opposite
sign. This is a clear signature of a MBS~\cite{Doru2012}.

The third row shows the spin of the YSR states. From the above data,
we have evidence that a topological phase transition (TPT) has taken
place between $J=2.1$ eV and $J=2.5$ eV , $J=2.3$ eV being near to the
closing of the gap.  The spin shows it unambiguously. The YSR bands  show opposite
spin polarizations for their particle and hole components. This is clearly
seen across the YSR gap. But the character has changed between $J=2.1$
eV and $J=2.5$ eV because the spin polarization is the opposite one. This
is a clear hallmark of a TPT~\cite{Mahdi2020}. The edge states show the
same spin polarization as corresponds to the MBS~\cite{Doru2012}.

The experimental data show that the gap is almost closed for the tetramer
$Cr_4$ $\sqrt{2}a-\langle110\rangle$ spin chain.  Closing the gap is a
necessary condition for a topological phase transition (TPT). 
Figure~\ref{figure4} clearly show that the
edge states for $J$ larger than 2.3 eV are indeed MBS, and that the
TPT takes place somewhere close to 2.3 eV.
The change of YSR band character through the TPT is clearly seen in the YSR
spin polarization~\cite{Mahdi2020}, indeed the spin inverts across the
transition.

\begin{figure*}
	\begin{center}
		\includegraphics[width=0.7\textwidth]{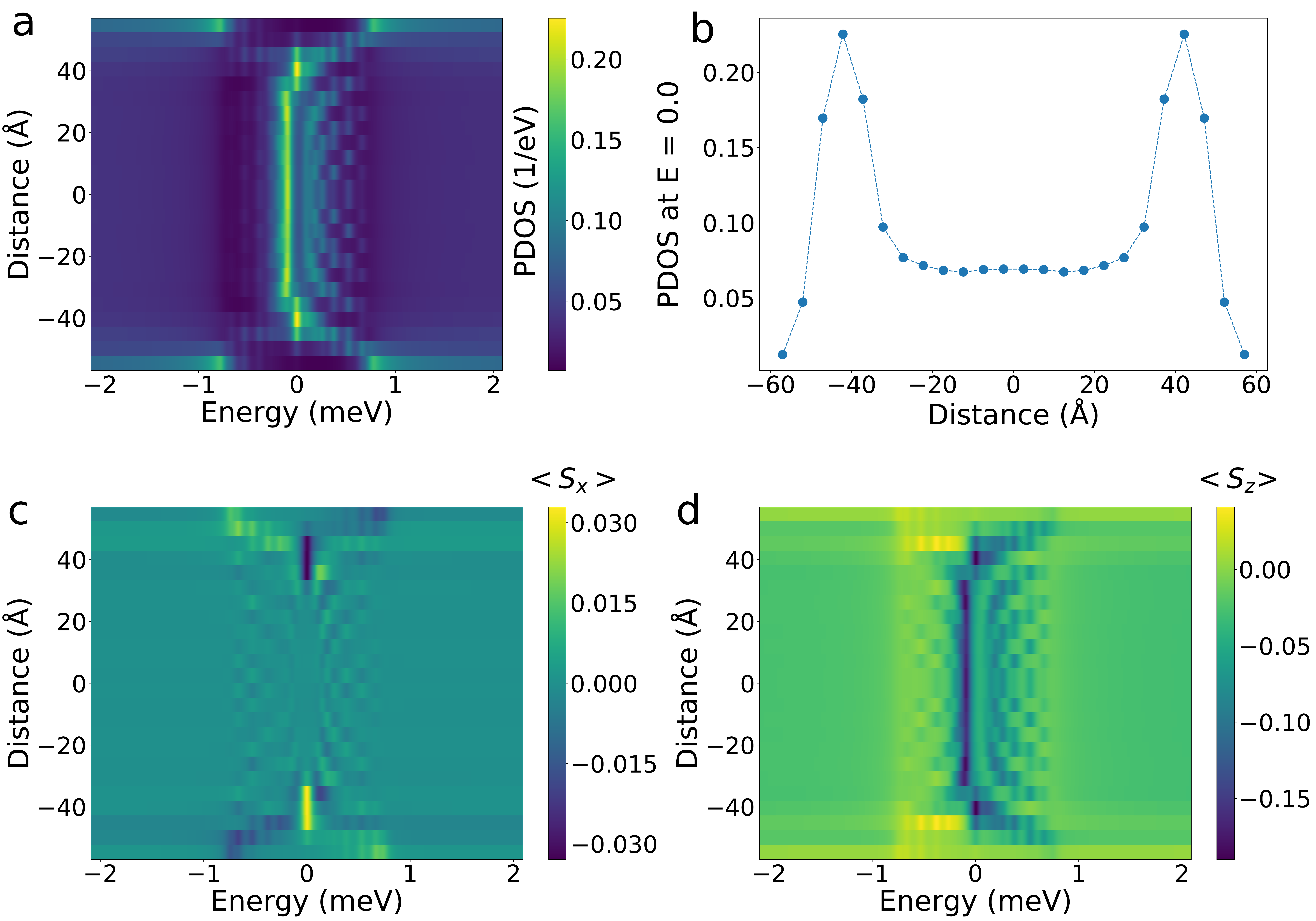}
	\end{center}
	\vspace{-0.2cm}
	\caption{
	Majorana bound states in a 20-atom $\sqrt{2}a-\langle110\rangle$
	Cr spin chain. (a) Color map plotting the PDOS as a function of energy and distance. 
	There is a clear state localized at the edges of the chain and at exactly zero
	energy. (b) PDOS along the 20-Cr chain, $x$-axis are distances
	in~\AA~along the Cr chain.
	The localization of the PDOS to the
	edges at zero energy spans the four Cr edge
	atoms, and the PDOS sharply fall beyond. The value of
	the PDOS between edges reduces as the chain length increases. 
	(c) Color map (dark: negative, light: positive) showing that the transversal spin density, $\langle S_x \rangle$, changes sign with
	edge, but (d) the out of plane spin-density component, $\langle S_z \rangle$, is the same for both edges.
	Moreover, these data can be correlated with a clear change of spin sign across the
	gap as the exchange-interaction value, $J$,  is increased
	that shows the closing and reopening of the gap into the
	topological phase. All
	these data signal the presence of a Majorana bound state in a 20-atom
	$\sqrt{2}a-\langle110\rangle$ Cr spin chain.
		\label{figure5} } 
\end{figure*}

Figure~\ref{figure5} shows the calculation of a $Cr_{20}$
$\sqrt{2}a-\langle110\rangle$ spin chain with  $J=2.5$ eV. A clear
spin-polarized edge state appears, with opposite transversal spin
components ($ \langle S_x \rangle$) on the chain edges showing that
indeed MBS are formed~\cite{Doru2012}.  The number of atoms in the
spin chain is decisive to clearly show  MBS. However, short chains
may suffice to prove that indeed the superconductor undergoes a TPT.

For a spin chain in the topological phase, the appearance
of the MBS needs a certain minimum chain size. This
is because the MBS have a certain extension and they overlap
for small chains. The consequence is that the zero-energy
state becomes localized in the center of the chain, and
it is difficult to identify the new superconducting phase
as topological.

\begin{figure*}[t]
	\includegraphics[width=1.0\textwidth]{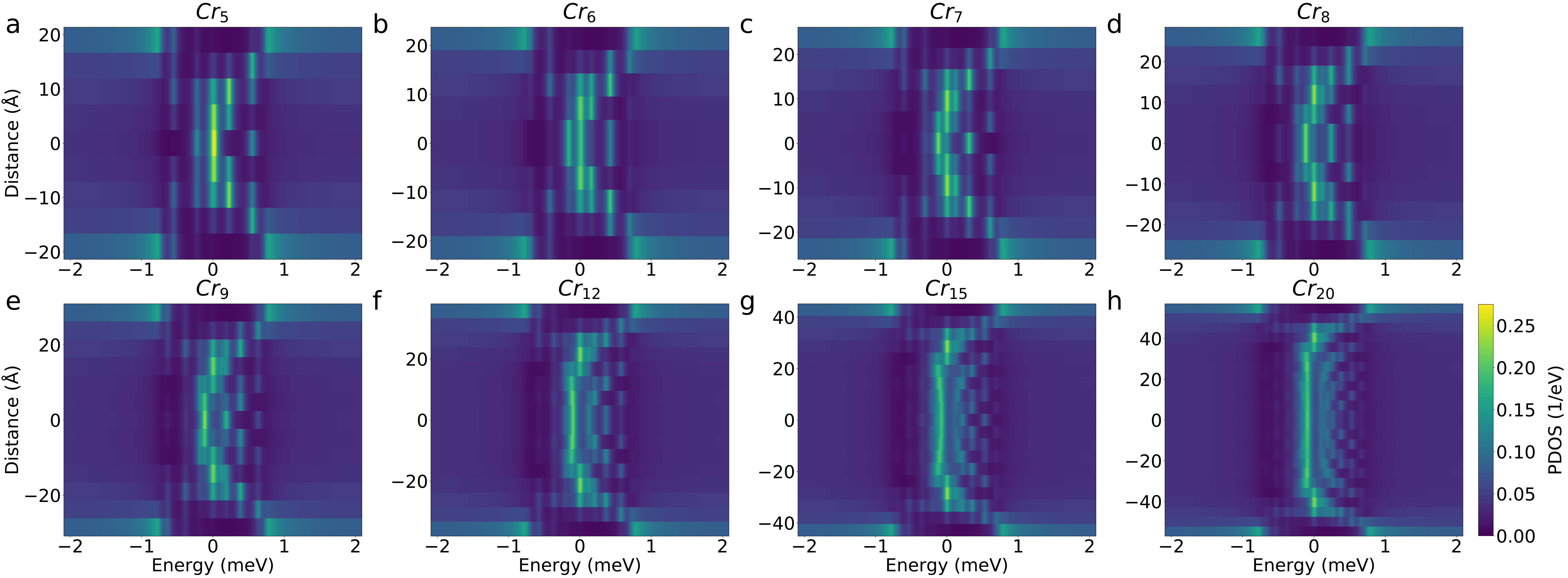}
	\vspace{-0.3cm}
	\caption{
		$Cr_n$ $\sqrt{2}a-\langle 110 \rangle$ chains
		with $n$ from 5 to 20, for $J=2.5$ eV such
		that the superconductor is in 
		the topological phase. The zero-energy
		state moves away from the center of the chain
		to the borders as the chain is increased in size. At
		fairly low numbers, 8 or even 7 atoms, the MBS
		become clear and a gap is formed at the center
		of the chain.
		\label{figure6} }
\end{figure*}

The behavior of MBS with the chain's length is shown in Fig.~\ref{figure6} 
for
 $Cr_n$ $\sqrt{2}a-\langle 110 \rangle$ chains
with $n$ from 5 to 20. The parameters are the above ones with $J=2.5$ eV that
correspond to the topological phase. In the case of the pentamer,  Fig.~\ref{figure6} (a) clearly shows a closed gap. 
We find a zero-energy state for $Cr_5$
that looks very similar to the experimental (and theoretical) one for $Cr_4$.
The zero-energy state is clearly localized in the center of the chain. As
the chain length is increased, the state localizes to the edges. At
the same time there is an excitation gap appearing in the center of the chain.
For $n=8$ atoms, it is already possible to clearly differentiate the
features of the well-formed MBS even though the chain is still small and the YSR states
present a strong discrete nature.
As the length is increased, a clear
MBS appears. 
These calculations imply that $Cr_n$ $\sqrt{2}a-\langle 110 \rangle$ chains
on \BiPd will clearly show MBS and topological features at fairly
small chains. Indeed, 20 atoms suffice to have an unambiguous topological
spin chain.

\section{Conclusion}

In summary, Cr$_n$ $\sqrt{2}a-\langle110\rangle$ spin chains on
\BiPd show a fast closing of the superconducting gap as the number
of atoms in the chain increases. As few as four Cr atoms suffice
to have in-gap states closing down the gap.  We show that an 8-atom
$\sqrt{2}a-\langle110\rangle$ chain may already display all features
of MBS.  Our study reveals that the $\sqrt{2}a-\langle110\rangle$ Cr spin
chain shows ferromagnetic alignment of its spins. The large magnetic
moment of Cr plus a sizable Rashba coupling of the \BiPd surface leads
to the topological phase transition. Increasing the distance between Cr
atoms leads to facile atom manipulation that translates in longer chains
of Cr atoms on \BiPd but at the cost of not reaching a topological phase.
Indeed, our measurements show a persistent gap rather constant with
chain length for Cr$_n$ $2a-\langle100\rangle$, showing that this type
of chains will not induce a topological phase transition on the \BiPd
superconductor.

\section*{Acknowledgement}
Financial support from the Spanish MICINN (projects RTI2018-097895-B-C44
and Excelencia EUR2020-112116), Eusko Jaurlaritza (project
PIBA\_2020\_1\_0017), JSPS KAKENHI (JP18K03531 and JP19K14651), and the Institute for Basic Science (grant IBS-R027-D1) is gratefully acknowledged. 

%
\bibliography{references}
\end{document}



\title{Supplementary Material for: Atomic Manipulation of In-gap States on the \BiPd Superconductor}

\author{Cristina Mier}
\altaffiliation{These authors contributed equally}
\affiliation{Centro de F{\'{\i}}sica de Materiales
	CFM/MPC (CSIC-UPV/EHU),  20018 Donostia-San Sebasti\'an, Spain}
\author{Jiyoon Hwang}
\altaffiliation{These authors contributed equally}
\affiliation{Center for Quantum Nanoscience (QNS), Institute for Basic Science (IBS), Seoul 03760, South Korea}
\affiliation{Department of Physics, Ewha Womans University, Seoul 03760, South Korea}
\author{Jinkyung Kim}
\affiliation{Center for Quantum Nanoscience (QNS), Institute for Basic Science (IBS), Seoul 03760, South Korea}
\affiliation{Department of Physics, Ewha Womans University, Seoul 03760, South Korea}
\author{Yujeong Bae}
\affiliation{Center for Quantum Nanoscience (QNS), Institute for Basic Science (IBS), Seoul 03760, South Korea}
\affiliation{Department of Physics, Ewha Womans University, Seoul 03760, South Korea}
\author{Fuyuki Nabeshima}
\affiliation{Department of Basic Science, University of Tokyo, Meguro, Tokyo 153-8902, Japan}
\author{Yoshinori Imai}
\affiliation{Department of Basic Science, University of Tokyo, Meguro, Tokyo 153-8902, Japan}
\affiliation{Department of Physics, Graduate School of Science, Tohoku University, Sendai, Miyagi 980-8578, Japan}
\author{Atsutaka Maeda}
\affiliation{Department of Basic Science, University of Tokyo, Meguro, Tokyo 153-8902, Japan}
\author{Nicol{\'a}s Lorente}
\email{nicolas.lorente@ehu.eus}
\affiliation{Centro de F{\'{\i}}sica de Materiales
	CFM/MPC (CSIC-UPV/EHU),  20018 Donostia-San Sebasti\'an, Spain}
\affiliation{Donostia International Physics Center (DIPC),  20018 Donostia-San Sebasti\'an, Spain}

\author{Andreas Heinrich}
\email{heinrich.andreas@qns.science}
\affiliation{Center for Quantum Nanoscience (QNS), Institute for Basic Science (IBS), Seoul 03760, South Korea}
\affiliation{Department of Physics, Ewha Womans University, Seoul 03760, South Korea}

\author{Deung-Jang Choi}
\email{djchoi@dipc.org}
\affiliation{Centro de F{\'{\i}}sica de Materiales
	CFM/MPC (CSIC-UPV/EHU),  20018 Donostia-San Sebasti\'an, Spain}
\affiliation{Donostia International Physics Center (DIPC),  20018 Donostia-San Sebasti\'an, Spain}
\affiliation{Ikerbasque, Basque Foundation for Science, 48013 Bilbao, Spain}


\date{\today}
\maketitle

\begin{figure*}[t]
\includegraphics[width=1.0\textwidth]{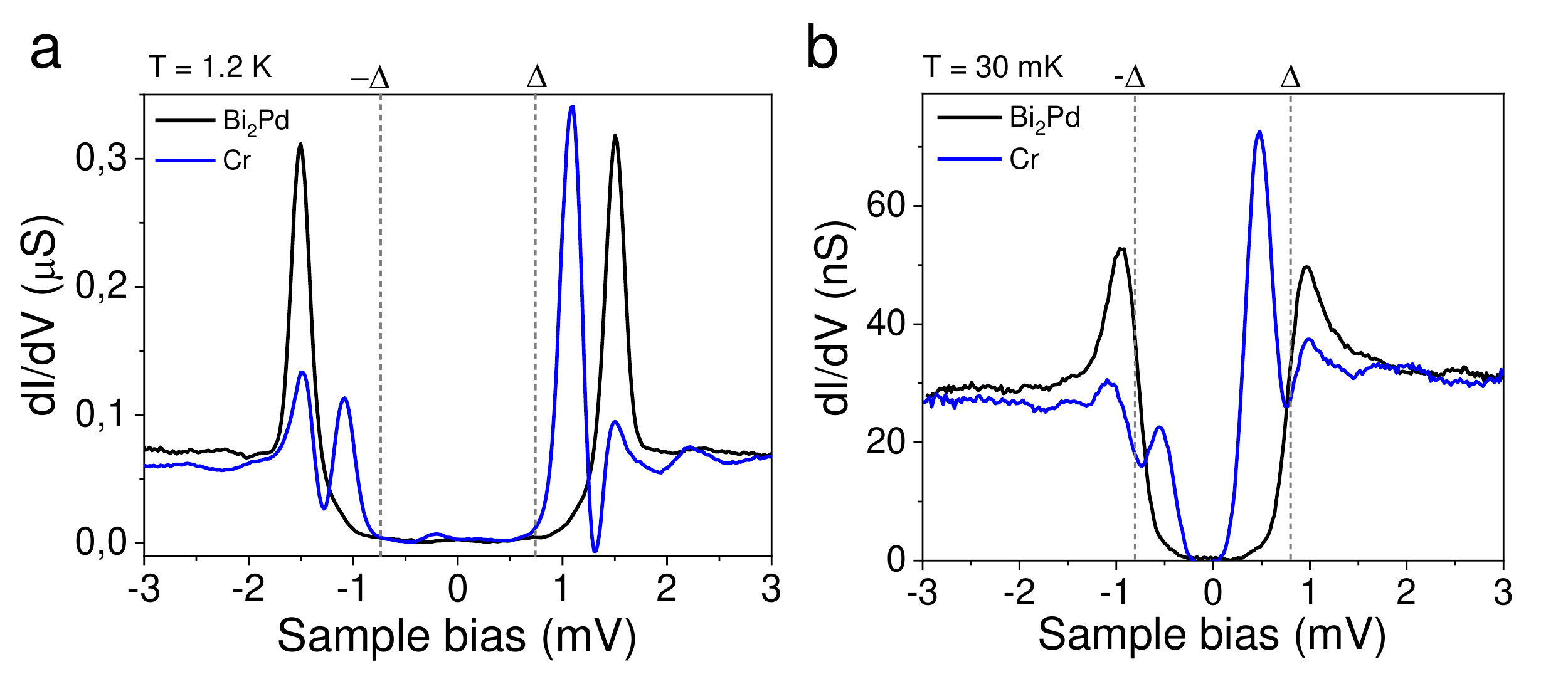} 
\vspace{-0.3cm}
\caption{
	Comparison of the differential conductance measured
	on a single Cr atom (blue curve) and on a bare
	\BiPd surface (black curve) with superconducting and
	non-superconducting tips.  The dI/dV spectra using (a)
	a superconducting (coated with Bi$_2$Pd) tip at T~=~1.2~K
        (adapted from Ref. \citenum{Choi_2018})
	and (b) a normal metallic PtIr tip at T~=~30~mK.
	\label{figure1} } 
\end{figure*}

\begin{figure*}[t]
\includegraphics[width=1.0\textwidth]{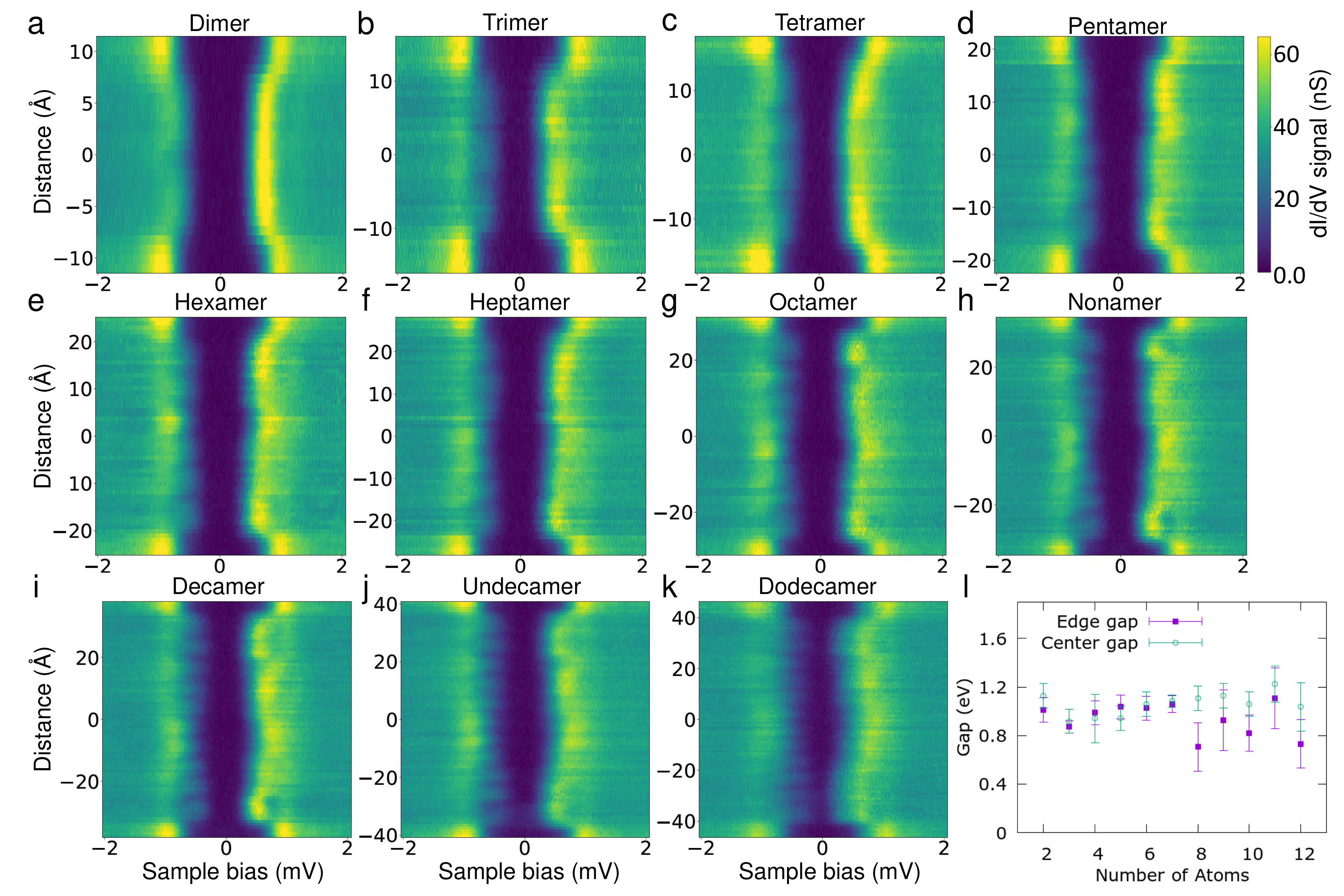} 
\vspace{-0.3cm}
\caption{
Differential conductance measured along for Cr$_n$ $2a-\langle100\rangle$ chains with $n=2 \cdots 12$.  
Figure 2 of the main text reproduces some of the pannels of this figure.
The gap has been obtained at an edge atom or at the center of
the spin chain. In both cases the gaps are constant within the experimental error.
        \label{figure2} } 
\end{figure*}

\section*{Comparison of the dI/dV spectra using metallic and superconducting tips}

We used a dilution fridge STM operating at $T=
30$ mK~\cite{kim2021spin}. The benefit of measuring at a very-low temperature leads
to a negligible thermal smearing granting a higher energy resolution in
a range of a few tens of $\mu$eV.  Measuring with a superconducting tip
increases the energy resolution due to the sharp quasi-particle peaks at
the tip side~\cite{Mier2021} but afterward, a deconvolution process is required to get
the density of states of the sample~\cite{Ruby_2015,Choi2017}.  Figure~\ref{figure1} shows
 the comparison
of the dI/dV spectra measured using a superconducting tip at 1.2 K and
a non-superconducting metallic tip at 30 mK.  We uses a metallic PtIr
tip that permitted us to use the $dI/dV$ as a direct measurement of the
density of states of the substrate without any deconvolution process.

\section*{Local spectra of the Cr$_n$ $2a-\langle100\rangle$ spin chains}

Figure~\ref{figure2} shows all the spectra obtained along the spin chains
formed by Cr$_n$ $2a-\langle100\rangle$ on \BiPd with $n=2, \cdots, 12$.
The evolution is very smooth and the features are very similar. Namely, the
in-gap states are very close to the superconductor's quasiparticle peaks, and
a clear, almost constant gap is maintained. Beyond $n=8$, the gap measured at the
edge slightly closes down, and an excited-state edge state starts
developing. However, as far as $n=12$ the gap at the edge stays constant
and there is no indication of any gap closing for Cr$_n$ $2a-\langle100\rangle$ 
on Bi$_2$Pd.
The gap has been evaluated by taking the distance between peaks in the second
derivative, $d^2I/dV^2$, after a 5-point smoothing of the experimental data.

\section*{Density functional theory calculations}

The DFT calculations were performed using
the VASP code~\cite{vasp}. The calculations performed
here extend the ones in Ref.~\cite{Choi_2018}. The \BiPd slab was optimized using the
Perdew-Burke-Ernzerhof (PBE) form of the generalised gradient
approximation (GGA)~\cite{PBE}, obtaining a bulk lattice parameter
$a=b=3.406$~\AA~and $c=13.011$~\AA~in good agreement with other
DFT calculations and the experimental value
of $3.36(8)$~\AA~and $12.97(2)$~\AA~given in Ref.~\cite{Herrera}.
The  surface calculations were performed for Bi-terminated slabs with
four Bi layers and two Pd ones. The surface unit cell was taken as a
$6\times4$ lattice, where two Cr atoms can be placed at $2 a$ without
interaction between dimers. The k-point sampling was $1\times 3 \times 1$.
The structures were relaxed until forces were smaller than 0.01
eV/\AA~for the three topmost layers and the Cr structures.

\bibliography{references}